# Ultrasonic Oscillatory Two-phase Flow in Microchannels

Zhaokuan Lu[1], Eric D. Dupuis[2], Viral K. Patel[3], Ayyoub M. Momen[4], Shima Shahab[*,1,2]

[1]Department of Mechanical Engineering, Virginia Polytechnic Institute and State University, Blacksburg, VA 24061, USA

[2]Department of Biomedical Engineering and Mechanics, Virginia Polytechnic Institute and State University, Blacksburg, VA 24061, USA

[3]Multifunctional Equipment Integration Group, Buildings and Transportation Science Division, Oak Ridge National Laboratory, Oak Ridge, TN 37831, USA

[4]Ultrasonic Technology Solutions, 10820 Murdock Dr, Suite 104, Knoxville, TN 37932, USA

*Keywords: Microchannel, Microfluidics, Ultrasonic vibration, Two-phase flow, Computational fluid dynamics, Drying technology*

## Abstract

Experimental and numerical investigations are performed to provide an assessment of the transport behavior of an ultrasonic oscillatory two-phase flow in a microchannel. The work is inspired by the flow observed in an innovative ultrasonic fabric drying device using a piezoelectric bimorph transducer with microchannels, where a water-air two-phase flow is transported by harmonically oscillating microchannels. The flow exhibits highly unsteady behavior as the water and air interact with each other during the vibration cycles, making it significantly different from the well-studied steady flow in microchannels. The computational fluid dynamics (CFD) modeling is realized by combing the turbulence Reynolds-averaged Navier-Stokes (RANS) $k-\omega$ model with the phase-field method to resolve the dynamics of the two-phase flow. The numerical results are qualitatively validated by the experiment. Through parametric studies, we specifically examined the effects of vibration conditions (i.e., frequency and amplitude), microchannel taper angle, and wall surface contact angle (i.e., wettability) on the flow rate through the microchannel. The results will advance the potential applications where oscillatory or general unsteady microchannel two-phase flows may be present.

## 1. Introduction

The rise of microfabrication techniques has significantly advanced the application of microchannels which can be classified as channel with hydraulic diameter ranges from 10 μm to 200 μm[1]. They have been favored in compact heat exchangers because of the large surface-to-volume ratio and small volume[2]. Microchannels are also frequently used in biomedical micro-electromechanical systems (MEMS) to imitate the actual biological structures such as capillary vessels[3] and to perform particle manipulations[4,5]. Other applications take advantage of the microchannel to transport fluids and particles in microfluidic devices[6]. Flows in the microchannel can be substantially different from their macroscopic counterparts. At the microscale, the surface tension force at the solid boundary can have a significant impact on flow behavior far exceeding the body force such as gravity due to the small channel width[7]. For the same reason, the flow is typically in the low Reynolds number regime, exhibiting pure laminar behaviors.

---

*Address all correspondence to this author. E-mail address: sshahab@vt.edu

Flows through microchannels have been well investigated, beginning with Hagen and Poiseuille's investigations in the 19th century[8]. These experimental and theoretical studies have generally been focused on steady-state flows to analytically describe the flow patterns of incompressible Newtonian fluids seen in long tubes of small diameter. However, a lapse in the investigation was found for oscillatory flows through microchannels. The oscillation greatly complicates the dynamics of the flow by increasing the inertial effects, and making it different from the ordinary microscale flow with low Reynolds number and steady behavior. Oscillatory flow in the microchannel can be found in valveless micropumps[9] and frequently studied both experimentally and numerically[10,11]. Application of the oscillatory microchannel flow can also be found in molecule manipulation[12,13] where cells in the microchannel are deformed and separated by pressure-driven oscillatory flows. A recent study[14] of a harmonic oscillatory microchannel flow up to 100 Hz reveals a much more complex velocity profile at high vibration frequency compared to low frequency. A greater challenge arises from studying ultrasonic oscillatory (>20 kHz) flows in microchannels where experimental techniques are rendered difficult. It can then be seen that computational fluid dynamics (CFD) and low-order analytical solutions[15] might be the few available methods for characterizing these flows. The behavior of the oscillatory flow in microchannels could be further complicated if more than one fluid is present (e.g., air-water two-phase flow). Turbulence could be induced by the perturbation from interfacial surface tension for already high-inertia oscillatory flow.

Two-phase flow in microchannels is an intensively studied topic due to its application in micro heat exchangers[16] where refrigerant liquid and vapor coexist in the microchannels. Efforts have been mostly dedicated to study the steady-state two-phase flow patterns, which are found to be different from the macroscopic two-phase channel flow[17]. Despite extensive experimental investigations[17–21], CFD simulation[22–26] methodologies have been frequently explored in this research area because of the difficulties in experimental measurement and fabricating microchannels with different geometry (e.g., diameter) and surface wettability properties, which significantly affect the two-phase flow behavior[27]. This paper attempts to bridge a knowledge gap in the understanding of the ultrasonic oscillatory two-phase flow transport behavior in microchannels. Despite the above-mentioned works on steady two-phase flow and oscillatory single-phase flow in microchannels, oscillatory two-phase flow in the ultrasonic range, which features strong unsteady phase interaction, has not been thoroughly investigated. In this work, experimental and numerical investigations are performed based on the configuration of a novel ultrasonic drying prototype where a water-air two-phase flow is transported by oscillating microchannels. The significant impact of this research is the development of effective numerical modeling techniques which allows for better visualization and understanding of the flow than experiment alone.

Recently, researchers at Oak Ridge National Laboratory investigated a novel concept for direct-contact ultrasonic fabric drying using vibrating piezoelectric (PZT, lead zirconate titanate) transducers with arrays of microchannels[28–30]. This drying technique is proven to be more efficient than traditional thermal drying[31]. In this process, a piezoelectric transducer, which comprises of a thin steel plate with annular piezoelectric bimorph rings, shown in Figure 1(a), is actuated by an input harmonic voltage in the ultrasonic frequency. For the bimorph structure under actuation, a voltage is applied across the top surface of the upper piezoelectric layer where it is connected in series to a terminal on the bottom surface of the lower piezoelectric layer. The stainless-steel substrate plate is considered a perfect conductor. The poling directions of the layers are away from the substrate, denoting a bending mode deformation of the bimorph; however, the highest piezoelectric coupling is in the thickness direction, giving evidence that thickness mode deformations dominate the displacement of the structure[32]. Actuation of the bimorph with a harmonic current signal produces the steel plate oscillating deformation, Figure 1(e).

    The plate transducer is shown in Figure 1(a) and (d). It has an outer diameter of 30 mm and thickness of 50 μm containing an array of microchannels (~5000) with an inlet diameter of 70 μm, Figure 1(b1) and (b2), and an outlet of 10 μm, Figure 1(c1) and (c2). Figure 1(e) shows the plate deformation profile at one of the resonant frequencies of the transducer, i.e., 107 kHz, where each dot represents a microchannel. The upward acceleration of the channels due to the plate deformation creates a pressure gradient, ejecting water on the plate through the channel outlet. As the pressure gradient is reversed due to subsequent deceleration, air below the transducer enters the channel and mixes with the water. On the upper surface of the transducer, ultrasonic vibration generates faraday instability on the free surface, ejecting droplets from the bulk liquid, referred to as atomization[33]. This device vibrates at an ultrasonic range (> 20 kHz), pushing the limits of our understanding of high-frequency two-phase flows through the microchannels. The drying rate is related to the flow rate through the channel and atomization among others, which are both investigated in this paper. A drying rate expression of this microchannel transducer is proposed and improved by Dupuis[34,35] with coefficients obtained from fitting the model to experimental data. However, the understanding of the working principle of this device and potential optimization for increasing the drying rate is hindered by the difficulties posed by the microscale experiment. Therefore, CFD simulation becomes a necessity to investigate this kind of problem.

    In this paper, we utilize the ultrasonic drying configuration to discuss the CFD simulation approach of the ultrasonic oscillatory two-phase flow in the microchannel. An axisymmetric fluid domain of a single microchannel will be used to characterize the two-phase fluid properties when the channel vibrates in the ultrasonic regime, Figure 1(f1) and Figure 1(f2). Effects of channel geometry, surface wettability, and excitation properties will be investigated for increasing the flow rate through the microchannel, which directly corresponds to the time needed to dry a wetted fabric in contact with the transducer, leading to a revolutionary new prototype for direct-contact ultrasonic drying. Not only can this prototype be used for other industrial drying purposes, but the simulation techniques proposed in this paper could also be easily migrated to other potential applications that include oscillatory two-phase flow in microchannels such as oscillation-driven fluids mixing and micropumps involving exchange with external fluids, or more general unsteady microchannel multiphase flows[36].

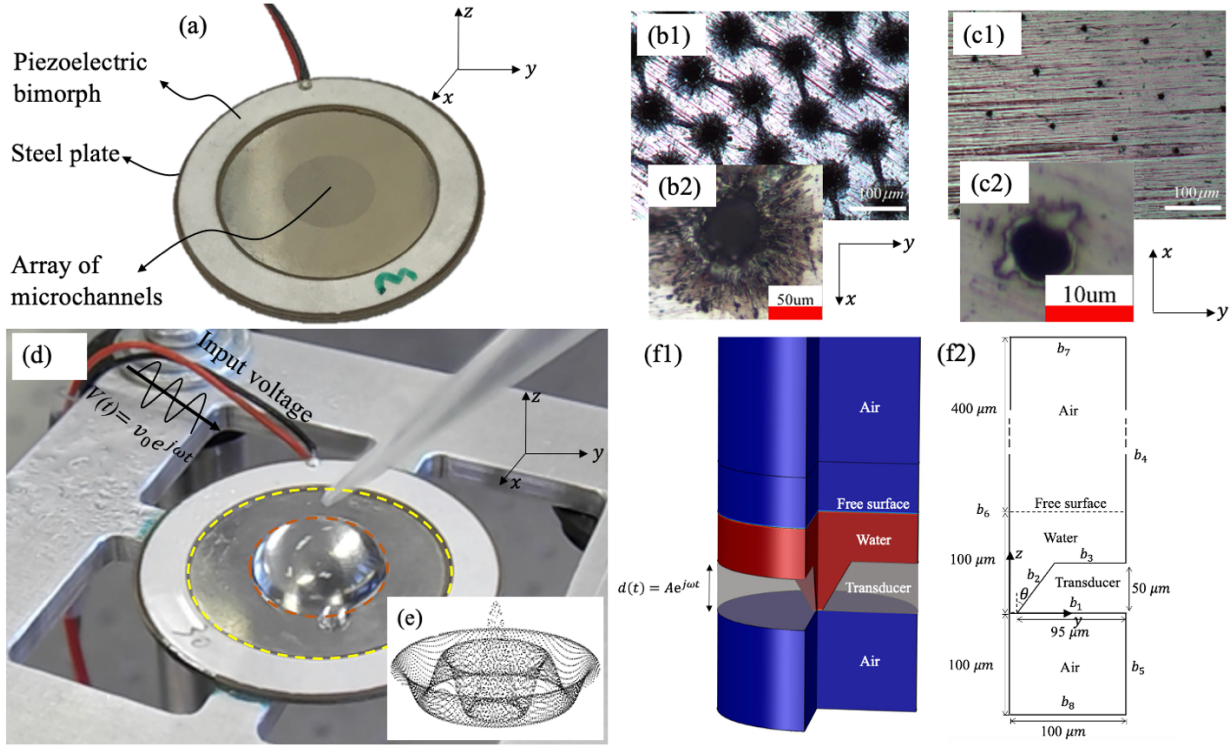

Figure 1 (a) Piezoelectric bimorph transducer with an array of microchannels, (b1) microchannel inlets on the upper side of the transducer, (b2) close-up of one of the inlets, (c1) microchannel outlets on the lower side of the transducer (c2), close-up of one of the outlets, (d) the transducer with a droplet placed on top of it before the vibration starts, (e) plate deformation profile at 107 kHz (one of the resonant frequencies), and (f1) the fluid domain of a single microchannel used to characterize the flow (a quarter of the domain is hidden for better visualization). The domain in red represents water, and that in blue represents air. The transducer is made transparent for better visualization of the fluid domain. The transducer performs harmonic vibration with displacement $d(t) = Ae^{j\omega t}$ where $A$ is the displacement amplitude and $\omega$ is the angular frequency. (f2) 2D axisymmetric fluid computational domain schematic (A cross-section of the 3D domain). The origin is put on the axisymmetric axis at the outlet.

## 2. CFD modeling

CFD models are used to simulate and characterize the two-phase flow in the microchannel. Since the fluid domain is axisymmetric, only a 2D cross-section needs to be modeled with boundary $b_6$ as the axis of symmetry (see the schematic graph shown in Figure 1(f2)). The water of 100 μm height is added to the transducer. The amount of water is small enough to observe the atomization within the first few oscillation cycles. The air domain extends 100 μm below the transducer and 400 μm above the free surface to observe the mist (water) ejection and atomization events. The channel wall (boundary $b_2$) tapers at angle $\theta$, which creates a pressure gradient to facilitate water ejection through the channel outlet. The structure is assumed rigid and thus not numerically modeled. Its vibration is imposed on the fluid field by specifying channel velocity on the fluid domain boundary $b_1$, $b_2$, and $b_3$ (non-slip boundary). For the right boundary $b_4$ and $b_5$, the symmetric condition is used, assuming the motion and water height of the adjacent channel are the same. The open boundary condition is applied at the upper and lower domain boundaries ($b_7$ and $b_8$) for free inward and outward water and air during the transducer vibration. The modeling and simulation are performed in Multiphysics software COMSOL®[37].

## 2.1 Fluid flow governing equations

The flow in the microchannel is modeled as a separated two-phase flow since a clear interface presents between the water and air, which are immiscible with each other. Therefore, the dynamics of both phases can be described by the Navier-Stokes equations. The flow is treated as being incompressible, assuming the water and air velocity both fall in the lower subsonic range and isothermal state is maintained in the channel due to the free heat exchange with the external atmosphere. Therefore, the density remains constant, and the energy equation and the equation of state (EOS) are excluded. Due to the absence of EOS, the water and air can be described by the same form of incompressible flow governing equations. Since the flow in the microchannel is driven by the oscillatory channel, and thus steady-state solution does not exist, unsteady continuity equation and momentum equation are used to describe the flow motion:

$$\nabla \cdot \boldsymbol{u} = 0 \tag{1}$$

$$\rho \frac{\partial \boldsymbol{u}}{\partial t} + \rho(\boldsymbol{u} \cdot \nabla)\boldsymbol{u} = \nabla \cdot [-p\boldsymbol{I} + \mu(\nabla \boldsymbol{u} + \nabla \boldsymbol{u}^T)] + \boldsymbol{F} \tag{2}$$

where $\boldsymbol{u}$ is the flow velocity vector, $\rho$ is the density, $\mu$ is the dynamic viscosity, $p$ is the flow pressure, and $\boldsymbol{F}$ is the body force vector. If the compressible effect has to be considered due to higher flow velocity, the equation of state for each phase along with the energy equation need to be incorporated in the governing equation to account for the strong coupling between the velocity, pressure, and temperature field, which is beyond the scope of this study. Although the regular single-phase flow found in the microchannel is laminar due to the small flow velocity and characteristic dimension (low Re number), increased flow inertia by the ultrasonic channel oscillation and perturbation from the phase interfacial stress could easily transform the flow into turbulence. To capture the turbulence effect, the revised RANS $k - \omega$ model proposed by Wilcox[38] is employed due to its robustness and excellent performance near the solid surface, whose motion has the largest impact on the flow behavior in the microchannel. In the RANS formulation, the flow velocity field and pressure become their averaged value denoted by $\boldsymbol{U}$ and $P$. Eq. 1 and Eq. 2 are reformulated as:

$$\nabla \cdot \boldsymbol{U} = 0 \tag{3}$$

$$\rho \frac{\partial \boldsymbol{U}}{\partial t} + \rho(\boldsymbol{U} \cdot \nabla)\boldsymbol{U} = -\nabla(P + \frac{2}{3}\rho k) + \nabla \cdot (\mu + \mu_T)(\nabla \boldsymbol{U} + (\nabla \boldsymbol{U})^T) + \boldsymbol{F} \tag{4}$$

where $k$ is the turbulence kinetic energy and $\mu_T$ is the eddy viscosity (a.k.a. turbulence viscosity), both introduced by Boussinesq approximation to describe the effect of turbulence fluctuation on the mean flow field. $\mu_T$ is calculated by a linear eddy viscosity model:

$$\mu_T = \rho \frac{k}{\omega} \tag{5}$$

where $\omega$ is the specific dissipation rate. Two additional transport equations of $k$ and $\omega$ are solved to close the fluid governing equation:

$$\rho \frac{\partial k}{\partial t} + \rho \boldsymbol{U} \cdot \nabla k = P_k - \rho \beta^* k\omega + \nabla((\mu + \sigma^* \mu_T)\nabla k) \tag{6}$$

$$\rho \frac{\partial \omega}{\partial t} + \rho U \cdot \nabla \omega = \alpha \frac{\omega}{k} P_k - \rho \beta \omega^2 + \nabla \cdot ((\mu + \sigma \mu_T) \nabla \omega) \tag{7}$$

The auxiliary variables and constants are defined as follows[37,38]:

$$\alpha = \frac{13}{25}, \beta = \beta_0 f_\beta, \beta^* = \beta_0^* f_{\beta^*}, \sigma = \frac{1}{2}, \sigma^* = \frac{1}{2}, \beta_0 = \frac{13}{125}, f_\beta = \frac{1+70\chi_\omega}{1+80\chi_\omega}$$

$$\chi_\omega = \left|\frac{\Omega_{ij}\Omega_{jk}S_{ki}}{(\beta_0^*\omega)^3}\right|, \beta_0^* = \frac{9}{100}, f_{\beta^*} = \begin{cases} 1, & \chi_k \leq 0 \\ \frac{1+680\chi_k^2}{1+400\chi_k^2}, & \chi_k > 0 \end{cases}, \chi_k = \frac{1}{\omega^3}(\nabla k \cdot \nabla \omega) \tag{8}$$

$$\Omega_{ij} = \frac{1}{2}\left(\frac{\partial \bar{u}_i}{\partial \bar{x}_j} - \frac{\partial \bar{u}_j}{\partial \bar{x}_i}\right), S_{ij} = \frac{1}{2}\left(\frac{\partial \bar{u}_i}{\partial \bar{x}_j} + \frac{\partial \bar{u}_j}{\partial \bar{x}_i}\right)$$

$$P_k = \mu_T(\nabla U : (\nabla U + (\nabla U)^T))$$

Due to the wide range of velocity scale caused by the unsteady nature of the flow, it is difficult to guarantee the dimensionless wall distance $y^+$ in the wall boundary layer to stay at a certain value. Therefore, using only the wall function which requires $y^+$ to fall into the logarithmic layer is not a feasible approach to model the near-wall flow. To deal with the constantly changing $y^+$, we use an automatic wall treatment proposed by Menter[39] which can resolve the flow down to the wall surface when $y^+$ is small enough (within viscous sublayer) and applies the wall function when $y^+$ is large. To accommodate the effect of channel motion on fluid domain deformation, Arbitrary Lagrange Eulerian (ALE) method is employed. The fluid mesh is smoothly deformed by the constraint imposed by the moving boundaries without changing the mesh topology (i.e., element connectivity).

## 2.2 Phase-field method

Since both fluids (i.e., water and air) share the same form of continuity and momentum equation, only one set of governing equations are needed to solve the dynamics of the two-phase flow. To determine the fluid properties such as density and viscosity at any point in the two-phase flow and incorporate the interfacial force, the phase interface needs to be tracked as the flow progresses. Interface tracking methods are generally classified as sharp and diffuse interface methods. Sharp interface methods treat the phase interface as a surface with no thickness (e.g., 2D surface in 3D configuration), which is reasonable since the interface thickness is typically negligible compared to the characteristic flow scale. The most common approach is to define a marker variable that is assigned a certain value in each fluid phase, such that the value is constant on the interface. Then the marker variable is propagated by the flow velocity using a convection equation. Popular two-phase flow schemes such as the level-set method, volume-of-fluid method, and front-tracking method all fall into this category[40]. Despite its easy implementation, the sharp interface method requires reconstruction of the interface in each time step by searching for the isosurface of the marker variable that represents the interface. This is not always trivial and even error-prone because the numerical inaccuracy during the marker variable convection may severely distort the interface geometry and thus leads to a wrong surface tension force being imposed on the fluid. In this work, we use the diffuse interface method which is also called the Phase Field Method (PFM). In PFM, the interface variables smoothly vary across a thin finite-width transition layer[41], which is based on an observation by van der Waals[42] that the phase interface is a mixing layer between the fluids driven by intermolecular diffusion and the interface is only in equilibrium if the total free energy on it diminishes to zero. Instead of the actual transition layer thickness, PFM uses a numerically acceptable value that is larger than the actual one. A phase variable $\phi$ is introduced to distinguish the different fluid phases by specifying a constant value in each bulk fluid (-1 for fluid 1 and 1 for fluid 2) and smoothly changed across the diffuse layer. The evolution of the phase-field variable $\phi$ is governed by the Cahn-Hilliard equation (convection-diffusion equation):

$$\frac{\partial \phi}{\partial t} + \boldsymbol{U} \cdot \nabla \phi = \nabla \cdot \frac{\gamma \lambda}{\epsilon^2} \nabla \psi \tag{9}$$

where $\psi$ is the free energy function defined as:

$$\psi = -\nabla \cdot \epsilon^2 \nabla \phi + (\phi^2 - 1)\phi \tag{10}$$

where $\boldsymbol{U}$ is the fluid velocity vector from the continuity and momentum equation serves to transport the phase variables with the flow, $\gamma$ is the mobility parameter which governs the diffusion time scale of the interface, $\lambda$ is the mixing energy density and $\epsilon$ is the capillary width which scales with the diffuse layer thickness. The finite-width phase interface is identified by the location where the variation of $\phi$ between limiting values of the two phases (-1 and 1) happens. The volume fraction of the fluid 1 is defined as:

$$V_{f1} = \min\left(\max\left(\frac{1-\phi}{2}, 0\right), 1\right) \tag{11}$$

with lower limit 0 and upper limit 1 and the volume fraction of the fluid 2 ($V_{f2}$) naturally becomes $1 - V_{f1}$. In this paper, we set water as fluid 1 and air as fluid 2. The fluid density and dynamic viscosity in the momentum equation (Eq. 4) can then be calculated by:

$$\rho = \rho_2 + (\rho_1 - \rho_2)V_{f1} \tag{12}$$

and:

$$\mu = \mu_2 + (\mu_1 - \mu_2)V_{f1} \tag{13}$$

where subscripts 1 and 2 represent the property of fluid 1 and 2, respectively. One of the benefits of PFM is that the phase interface is not required to be explicitly reconstructed to calculate the surface tension force as in the sharp-interface method. Instead, the surface tension force, $\boldsymbol{F}_{st}$, is simply obtained by:

$$\boldsymbol{F}_{st} = G \nabla \phi \tag{14}$$

where $G$ is the chemical potential in the fluid defined as a function of the free energy $\psi$:

$$G = \frac{\lambda \psi}{\epsilon^2} \tag{15}$$

Since this relationship is valid throughout the fluid, $\boldsymbol{F}_{st}$ is calculated at all fluid nodes and directly incorporated into the body force term $\boldsymbol{F}$ in the momentum equation (Eq. 4). The surface tension treatment in PFM is better than the sharp interface method because the interface reconstruction is not required, and the surface tension force is more physically motivated. To solve the Cahn-Hilliard equation (Eq. 9 and Eq. 10), boundary conditions are specified as:

$$\boldsymbol{n} \cdot \epsilon^2 \nabla \phi = \epsilon^2 \cos(\theta_w)|\nabla \phi| \tag{16}$$

for phase-field variable $\phi$ to enforce the contact angle $\theta_w$, and

$$\boldsymbol{n} \cdot \frac{\gamma\lambda}{\epsilon^2} \nabla\psi = 0 \tag{17}$$

for free energy $\psi$ as a zero-flux condition. $\theta_w$ is the contact angle (between phase interface and solid surface on the fluid 1 side) which can be obtained from Young's equation:

$$\sigma_{12}\cos(\theta_w) + \sigma_{s2} = \sigma_{s1} \tag{18}$$

where $\sigma_{s1}$ and $\sigma_{s2}$ are the surface tension on the fluid-solid interface for fluid 1 and fluid 2, and $\sigma_{12}$ is the surface tension on the interface of fluid 1 and fluid 2. In PFM, the motion of the fluid phase interface on the solid surface (contact line) is naturally taken care of by the diffusion of free energy (RHS of Eq. 9)[43]. On the other hand, an equivalent mechanism does not exist for the sharp interface method where a Navier-slip boundary condition has to be enforced to incorporate the contact line motion on the solid surface[44]. The slip length used to prescribe the Navier-slip boundary condition usually has to be determined experimentally.

In addition to the better modeling of surface tension and contact line motion, PFM guarantees energy and volume conservation[45,46] which are not achieved by some of the sharp-interface methods (e.g., level-set method). Although the diffuse layer thickness may not be made as small as the physical size of the phase interface[47] such that more numerical diffusion is introduced, a sharp-interface limit could be approached asymptotically as the capillary width $\epsilon$ (which scales with interface thickness) is reduced[43].

## 2.3 Numerical methods for the two-phase flow

The numerical discretization methods to solve the coupled fluid governing equations and phase-field equations in the spatial and temporal domain are described. For the spatial discretization, the Finite Element Method (FEM) is employed with the mesh comprised of first-order triangular element for the bulk fluid domain and six prism element layers close to the wall to take care of the boundary layer flow. Adaptive meshing has been a popular approach to simulate two-phase flow problem. Mesh refinement is only performed around the phase interface so that the node number is confined within a reasonable value. However, adaptive meshing is not particularly attractive in this study. Given the highly transient nature of the flow, the base-level mesh, run before the mesh refinement to locate the interface, has to be very fine and updated at short intervals to correctly capture the interface motion, which already consumes a large amount of computation even before the adaptively refined mesh is run. In this study, a fixed mesh (non-adaptive) with an almost uniform element size will be used because the phase interface may reach any part of the domain as the flow progresses. Capillary width $\epsilon$, which regulates the thickness of the diffuse layer, is set to be the maximum element size to guarantee sufficient grid points within the transition layer throughout the fluid mesh, following the recommendation by Yue[48]. This is another reason to make the element size more or less uniform (i.e., maximum element size close to minimum element size) since $\epsilon$ should be as small as possible to avoid excessive diffusion and keep the interface sharp. A discussion about the mesh independence study can be found in Appendix A. For temporal discretization, the implicit backward Euler method is used due to the fact that it has high numerical stability, and explicit formulations are not readily available for nonlinear partial differential equations involved in this problem.

## 3. Results and discussion: experiment and modeling

## 3.1 Experiments

To characterize the vibrational responses and resonant frequencies of the piezoelectric bimorph transducer (shown in Figure 1(a) and (d)) to electrical stimulus, the frequency-response function (FRF) of the transducer plate to an AC voltage input was found, experimentally (Figure 2). A sine wave was generated with National Instruments (NI) Signal Express and amplified with a Krohn-Hite model 7500 amplifier. The frequency of the wave was increased from 1 Hz to 200 kHz. A reflective tape was placed at the geometric center of the transducer where a single-point laser doppler vibrometer (LDV; Polytec OFV 5000/505) measured the velocity of oscillation at the center of the steel plate, from which the acceleration was calculated for the harmonic response of the plate. The experimental configuration is shown in Figure 1(d) and Figure 3(a). The transducer was clamped to a fixture which was then fixed on an optical table. The frequency sweep was performed at 30 V input voltage and the resonant frequencies are represented by the individual peaks in Figure 2. The obtained experimental FRF was then validated by FEM analysis (detailed information can be found in Dupuis[34]) while the linear elastic structural response assumption was held. In Figure 2, at frequencies higher than 107 kHz, the agreement between the analytical and experimental results starts to break down due to structural nonlinearities. As a result, the studied frequency domain in this work is chosen within the range from 20 kHz (i.e., ultrasound lower limit) to 107 kHz which is shown by dashed lines in Figure 2. Having identified the resonant frequencies of the transducer, we aimed to characterize the flow rate of water during the drying process which is contributed by both the water flow through the microchannels and atomization. Two volumes of water (100 and 300 µL) were added on the transducer via a micro pipette, Figure 1(d) and Figure 3(a), covering the span of microchannels. In this experiment, the bimorph transducer was actuated by a 30V sinusoidal input voltage at 107 kHz, where the flow rate is available from the manufacturer. The time, until the volumes of water were dispersed, was measured with a stopwatch and the flow rate of water was calculated by the volume of water divided by the total drying time (5 seconds). Figure 3 (a) to (f) show the snapshot of the drying process, including the ejection of water in the form of mist and atomization. It can be seen that the majority of the fluid was ejected through the microchannels as a mist jet in the first few seconds. The atomization only happened at the last second of the drying process when the water was almost gone (t=4 s), Figure 3(e). The late onset of the atomization is because the free surface instability is not excited when the water height is high since the microchannel allows free exchange of fluid with the outside and thus damps the oscillatory motion at the free surface. As most water is drained through the microchannels and water height becomes small, the remaining water left on the upper surface of the transducer is subject to strong vibration excitation, which eventually leads to atomization.

Following this, experiments with voltage amplitudes of 30 V to 60 V in increments of 5 V were carried out for comparisons of measurements of the water flow rates, where each water volume was tested three times and the average of the measured times taken was recorded. The flow rate of water for 100 and 300 µL water columns at different input voltage is plotted in Figure 4. The manufacturer reported 3 mL/min at 30 V amplitude, which is close to that of the experimental measurements. In addition to the flow rate of water, acceleration amplitude was sampled at several points on the transducer using the LDV and the average acceleration is plotted along with the flow rate in Figure 5. Fitting the data points in Figure 5, a linear relationship between the acceleration and flow rate of water is identified. It can be seen that the intercept of the fitted function is negative, meaning the mist ejection and atomization do not happen until a certain threshold vibration amplitude. This is because there exists a minimum amount of inertia the water needs in order to overcome the initial surface tension of the water-air interface.

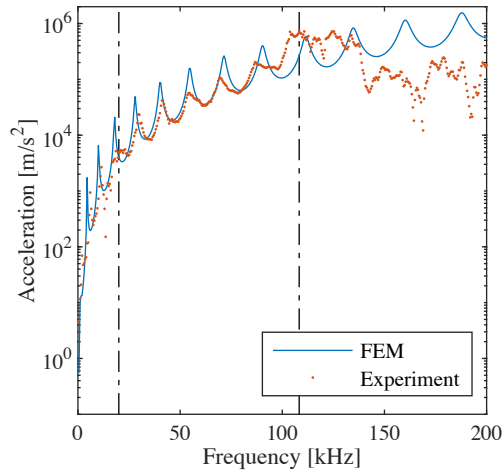

Figure 2 Experimental and numerical (FEM) acceleration FRFs. The data measured and calculated at the geometric center of the vibrating steel plate.

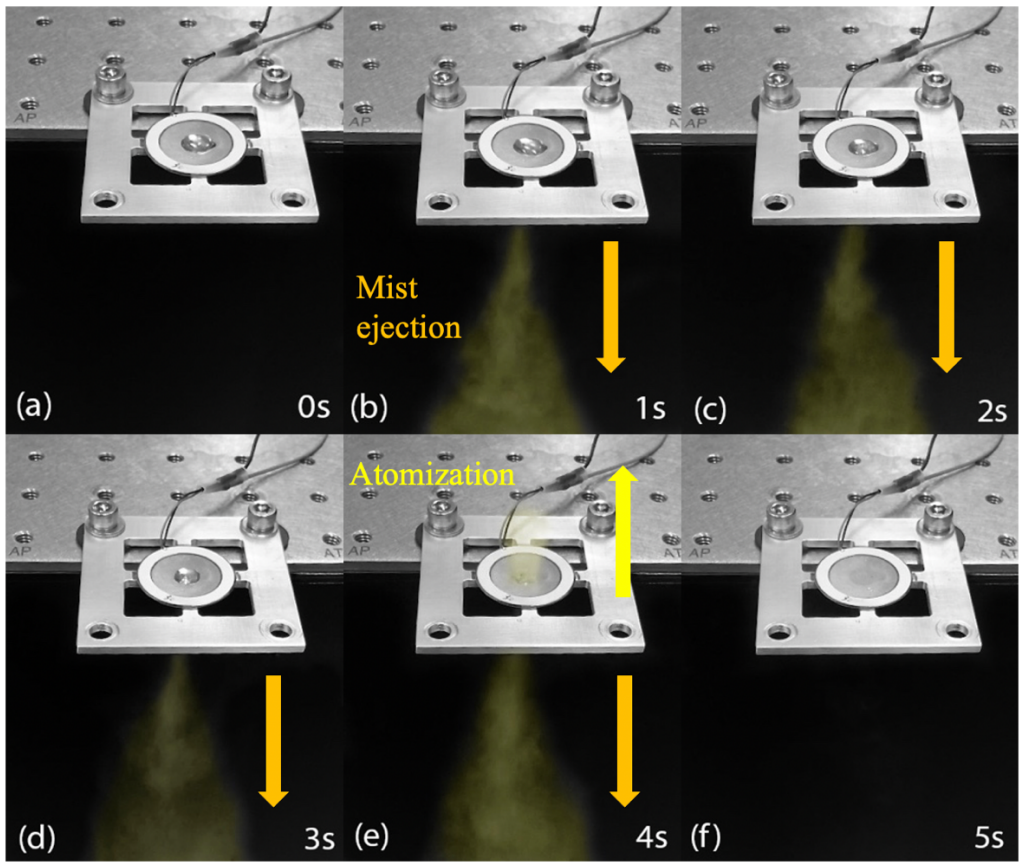

Figure 3 Snapshot of the drying process from 0~5s includes the mist ejection and atomization (graphically manipulated for better visualization).

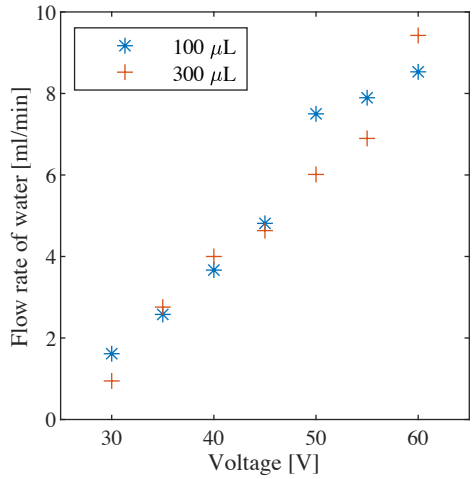

Figure 4 Experimental flow rate of water measured for different harmonic actuation voltages.

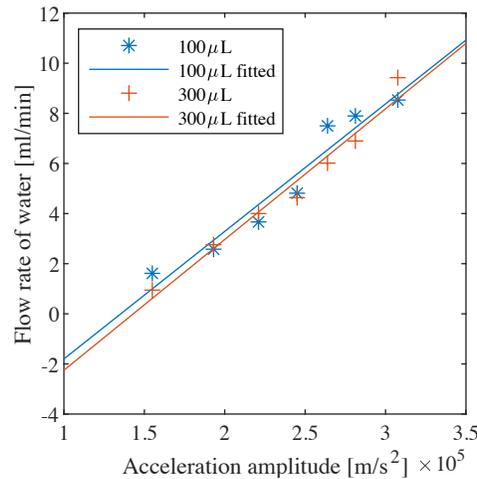

Figure 5 Experimental relation between the acceleration amplitude and the flow rate of water.

## 3.2 Numerical modeling

The experimental observations made in the previous section are used to validate the numerical method. To represent the physics of the oscillatory flow in the microchannels, only one microchannel is simulated to reproduce and characterize the flow phenomenon as well as the flow rate of water qualitatively, rather than quantitatively. This has an added benefit of avoiding impractical computations required to simulate all the channels (~5000) with different vibration conditions. Instead of the average vibration amplitude measured in the experiment, a series of amplitude that gives stable water ejection (that turns out to be higher than the average acceleration) is used. The reason is that the highest vibration amplitude that contributes the most to the total flow rate may not be captured by the sampled points used to quantify the average acceleration. The displacement amplitude in the simulation ranges from 25~50 μm in increments of 5 μm, and the frequency is fixed at 107 kHz as in the experiment. In the first quarter cycle of the vibration, we use a smoothed Heaviside function that is second-order differentiable to replace the harmonic vibration displacement profile, Figure 6(a), so that the boundary velocity starts from zero with zero first derivative instead of the velocity amplitude as shown in Figure 6(b). This both reflects the real situation and avoids numerical singularities.

Since the water contact angle, which depends on microfabrication details (e.g., material, surface roughness), is not known from the manufacturer, we assume a 90° contact angle on the channel surface, which is neither hydrophobic nor hydrophilic. The channel taper angle $\theta$ in Figure 1(f2) is set to 30°, consistent with the experimental setup. The initial water height is set at 100 µm, small enough to allow for atomization to be observed within the first few vibration cycles, rather than hundreds of thousands in the experiment (428k cycles performed when atomization happens at 4s, Figure 3(e)). Due to the dominance of water ejection in the drying process, the volumetric flow rate of water at the channel outlet ($Q_1$) is used to represent the total water flow rate. $Q_1$ is integrated in a polar coordinate on the channel outlet plane:

$$Q_1 = \iint_R w_1(r) dA = \int_0^{2\pi} \int_0^{r_{out}} w_1(r) \, r dr d\theta = 2\pi \int_0^{r_{out}} w_1(r) \, r dr \tag{19}$$

where $r$ is the radial coordinate integrated from 0 to the outlet radius $r_{out}$, and $\theta$ is the angular coordinate integrated from 0 to $2\pi$. Here $w_1(r)$ is the vertical velocity of water, depending only on $r$ due to the axisymmetric assumption:

$$w_1(r) = w(r) V_{f1}(r) \tag{20}$$

where $w(r)$ is the vertical two-phase flow velocity (positive upwards) and $V_{f1}$ is the volume fraction of water. Negative $Q_1$ represents the water going out of the channel through the outlet and positive $Q_1$ stands for the water going into the channel. The flow rate of water at the channel outlet for the 50 µm displacement amplitude case is plotted in Figure 6(c) along with the contour plot of water volume fraction, Figure 7, sampled at the time denoted by asterisks in the first quarter vibration cycle (shown in Figure 6(a)~Figure 6(c)). As shown in the volume fraction plot, Figure 7, and flow rate time history, Figure 6(c), water ejection happens almost instantly, driven by the pressure gradient created by the positive pressure in the channel and negative pressure below the channel as the channel accelerates upwards. The maximum flow rate is reached shortly after the ejection, around the time the maximum acceleration is reached, and the pressure gradient is the highest, Figure 7(b). As the highest velocity is reached (zero acceleration) and the channel is about to decelerate, the flow rate becomes almost zero, Figure 7(c). As the deceleration begins, the pressure gradient is reversed, which allows the air to enter the channel and pushes the water back, Figure 7(d) and Figure 7(e). The ejected water shown in the simulation will eventually break up into small droplets as it travels further downstream, which agrees with the observation in Figure 3 where the clear mist is observed below the transducer soon after the vibration starts. However, a close examination of the velocity field suggests that the air velocity at the outlet becomes supersonic soon after it enters the channel at this vibration condition, meaning the flow becomes compressible and cannot be properly solved by the current incompressible solver. As a result, the maximum water flow rate at the outlet, which happens before the air is entrained, is used for an estimation of the flow rate during the drying process. The normalized flow rate of simulation and experimental results are plotted in Figure 8 with respect to the normalized acceleration amplitude. Linear curve-fitting on the discrete data indicates the linear relationship between the acceleration amplitude and flow rate found in the experimental result is reproduced by numerical simulation. The coefficient of determination $R^2$ of the fitted functions is noted in the legend of Figure 8.

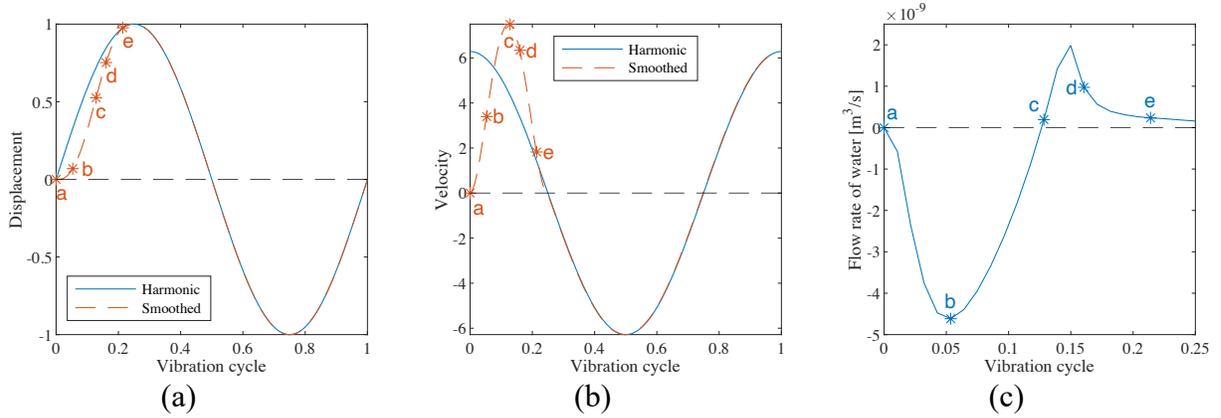

Figure 6 (a) Normalized harmonic vibration displacement profile, and (b) its corresponding velocity profile, and (c) the flow rate of water in the first quarter vibration cycle (107 kHz). A horizontal dashed line at zero is added to distinguish the direction of the flow. The sampled points are marked by asterisks and the corresponding letters in Figure 7.

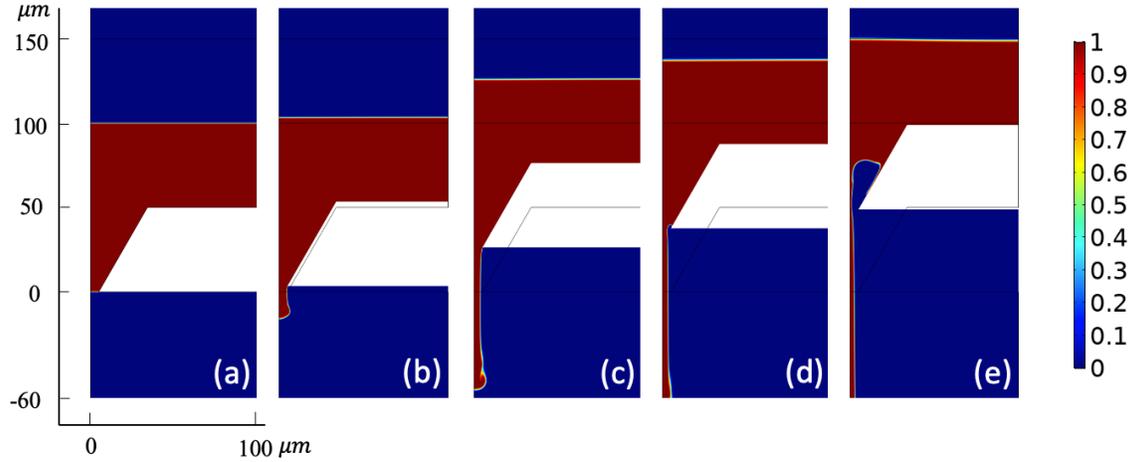

Figure 7 Volume fraction of water contour plot at different sample times defined in Figure 6 (107 kHz).

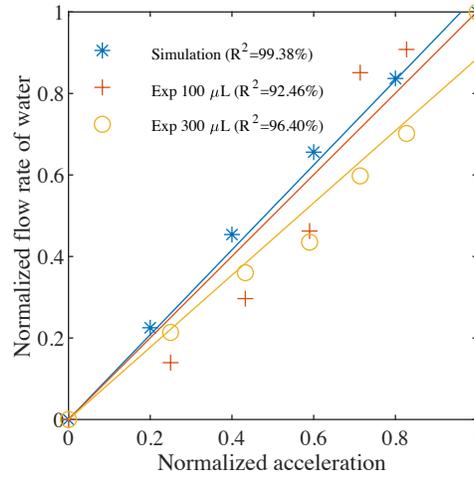

Figure 8 Experimental and calculated normalized flow rate of water at 107 kHz versus normalized acceleration. The axis normalization is performed by subtracting the minimum value and dividing by the maximum value to set the range from 0 and 1.

    To simulate the flow behavior in the microchannel for a longer duration with the current incompressible flow solver, the frequency is decreased to 29.5 kHz which is the first resonance frequency in the ultrasonic vibration range. In this configuration, the air velocity falls safely in the incompressible regime for the vibration amplitudes assessed in the previous simulation. The water flow rate at the outlet for the 50 μm vibration amplitude case and sampled volume fraction of water contour plots in a full vibration cycle are shown in Figure 9(c) and Figure 10, with the corresponding vibration profiles shown in Figure 9(a) and (b). It can be seen that the flow rate pattern in the first quarter cycle, Figure 10(a)~Figure 10(e), is similar to the 107 kHz case, Figure 7. As the vibration continues, the air keeps entering the channel by the pressure gradient created by the channel downward acceleration in the second quarter cycle, Figure 10(f). Then, the reversed pressure gradient in the third and fourth quarter cycle (due to upward acceleration) begins to pull the fluid towards the channel outlet, Figure 10(g), to the extent that a secondary water ejection takes place in the fourth quarter cycle as the channel restores its equilibrium position, Figure 10(h)~Figure 10(j). By the end of the first cycle, the free surface has become unstable, Figure 10(j). To examine if the linear relationship between the acceleration amplitude and flow rate persists at this frequency, 29.5 kHz, normalized maximum and average flow rate in the first vibration cycle are used to quantify the flow rate and plotted against normalized acceleration. Again, a linear relationship between the acceleration amplitude and flow rate is found, Figure 11, as suggested by the high $R^2$ (shown in legend) of the fitted linear functions.

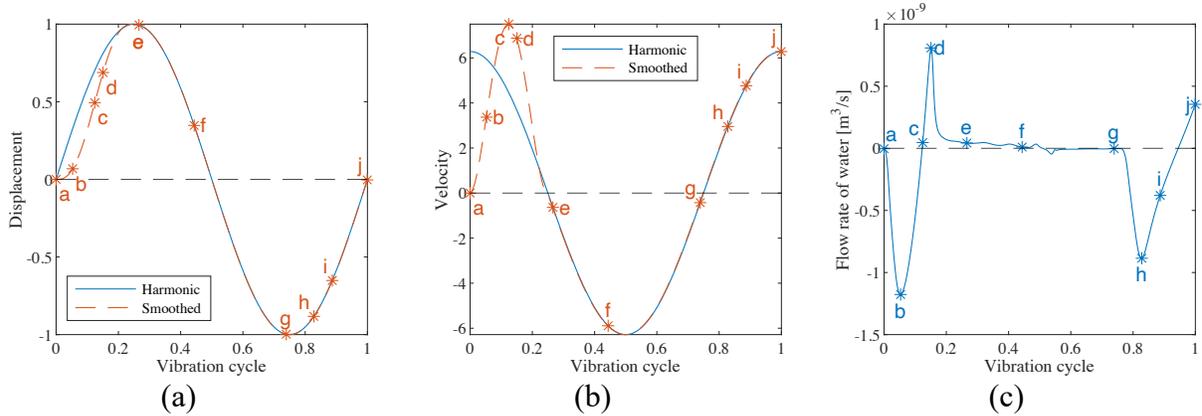

Figure 9 Normalized harmonic vibration displacement profile, and (b) its corresponding velocity profile, and (c) the flow rate of water in the first quarter vibration cycle (29.5 kHz). A horizontal line at zero is added to distinguish the direction of the flow. The sampled points are marked by asterisks and the corresponding letters in Figure 10.

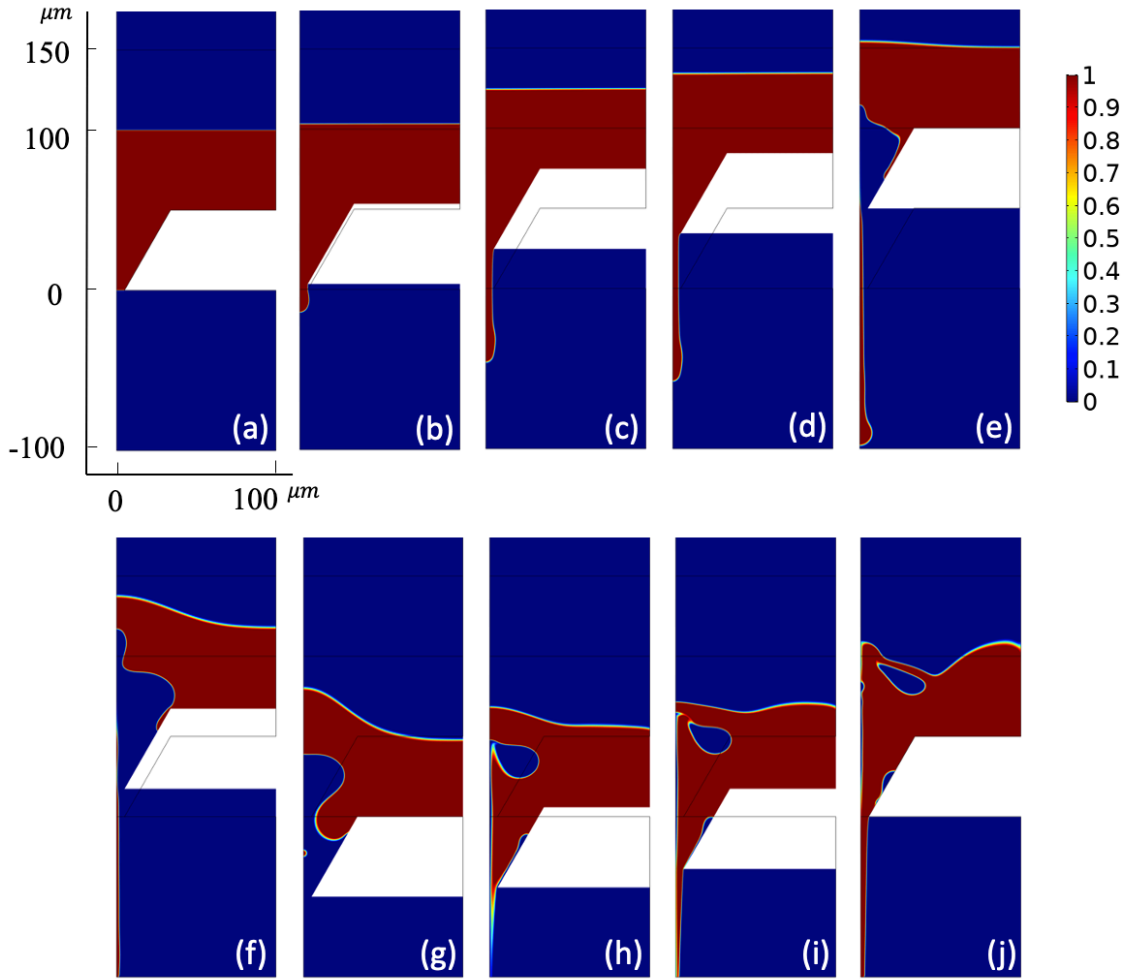

Figure 10 Volume fraction of water contour plot at different sample times, defined in Figure 9, in the first vibration cycle (29.5 kHz).

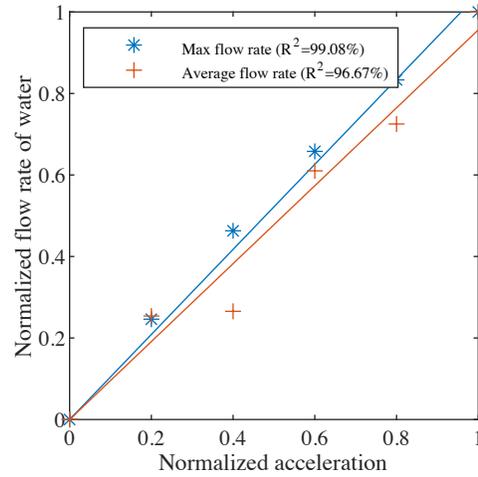

Figure 11 The normalized flow rate of water versus normalized acceleration at 29.5 kHz. The axis normalization is performed by subtracting the minimum value and dividing by the maximum value to set the range from 0 and 1.

    To visualize the atomization, the volume fraction of water contour plots are also depicted in the second vibration cycle, Figure 12, with the sampled time marked in the channel displacement profile, Figure 13(a). The flow rate of water at 150 µm, the highest position the free surface would reach if there is no atomization, is plotted in Figure 13(b) to evaluate the motion of flow as the atomization takes place. The results suggest that faraday instability happens at the free surface, Figure 12(a), and conical spikes are developed and pushed upwards away from the free surface as the instability escalates, Figure 12(b)~ Figure 12(d), in the second vibration cycle. When the neck of the spikes become thin enough as the spike moves further away, a pinch-off would break the bulbous tip from the cone and form a droplet, which leads to atomization on the upper surface of the transducer as shown in the experimental snapshot, Figure 3(e). Since the entrained air pushes the water upwards, the conical spike is higher above the channel, the left side of Figure 12(a)~(d), which indicates that the microchannel can enhance the atomization. It is also important to notice that the atomization happens much later than the water ejection, which agrees with the experimental observation, Figure 3. Although the flow rate of water is much larger than the outlet flow rate in the first vibration cycle, atomization only happens at a very late stage thus only a constitutes a small portion of the total water flow rate in a more realistic setup with much larger water height such as that in the experiment.

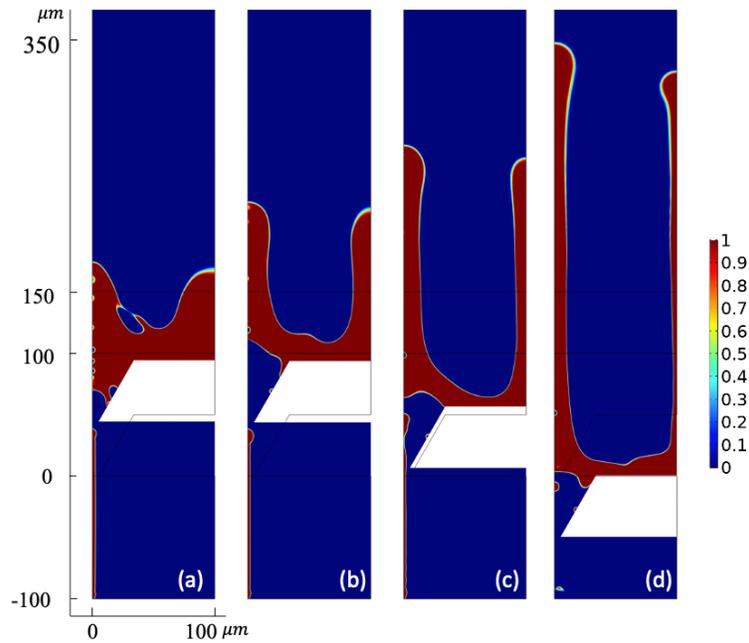

Figure 12 Volume fraction of water contour plot at different sample times in the second vibration cycle (29.5 kHz).

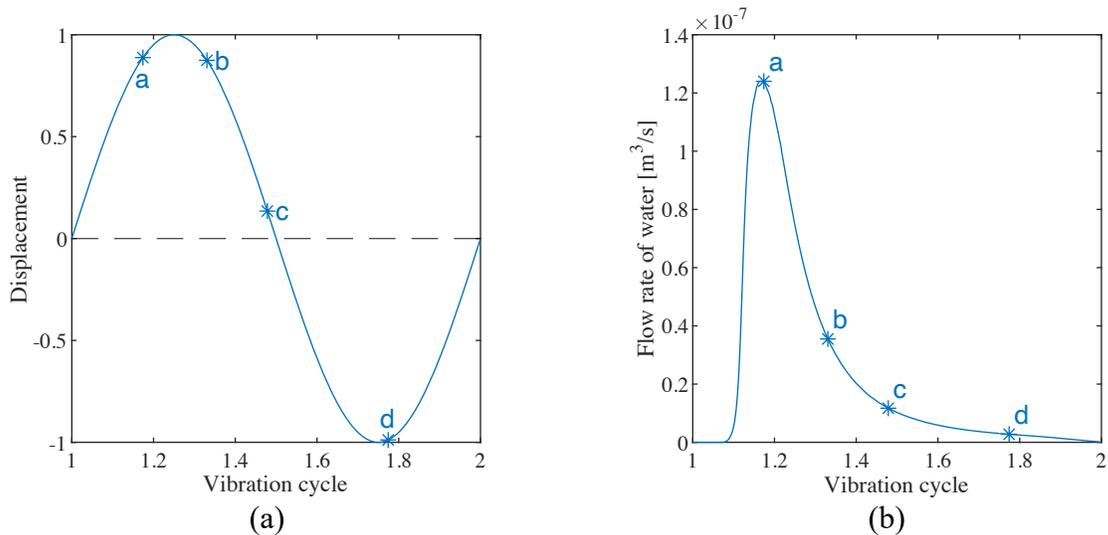

Figure 13 (a) Normalized displacement profile in the second vibration cycle, and (b) flow rate of water at 150 μm calculated by Eq. 19 with $r_{out}$ replaced by the width of the fluid domain (100 μm). Sampled points are marked by asterisks and corresponding letters in Figure 12.

## 4. Parametric study

In this section, parametric studies that are hindered by the lack of available experimental resources are performed by numerical simulation. Specifically, the effects of channel taper angle, vibration condition, and channel surface contact angle on the outlet water flow rate are assessed, and configurations for the highest flow rate are proposed. Water flow rate due to the atomization is neglected since it only constitutes a small portion of the total flow rate.

## 4.1 Effects of channel taper angle

As mentioned in Section 3, outlet water flow is driven by the pressure gradient created by the positive pressure of the fluid inside and the negative pressure of the fluid below the channel as it accelerates upwards during a harmonic vibration cycle. As the channel taper angle $\theta$, illustrated in Figure 1(f2), increases, larger normal boundary velocity is imposed on the fluid as the channel moves upwards, leading to a different pressure profile within the channel. To identify the optimal taper angle in terms of maximizing the water flow rate, taper angles from 0 to 60° in increments of 10° are simulated. The vibration frequency and displacement amplitude are set at 29.5 kHz and 50 μm. The outlet flow rate in the first vibration cycle plotted in Figure 14 is used to compare the behavior of different taper angles. The average flow rate (calculated by the total flow volume divided by the elapsed time) in the first vibration cycle is plotted against the taper angle in Figure 15, which suggests a 20° channel taper angle achieves the highest flow rate out of the microchannel. To explain the behavior of different taper angles, we estimate the pressure gradient that drives the flow out of the channel by the maximum difference of the average pressure on the boundary $b_1$ and $b_2$ shown in Figure 1(f2). The pressure difference plotted in Figure 16 exhibits an asymptotic behavior as the taper angle increases. Despite the larger pressure difference at a larger taper angle, the flow rate does not follow the same trend, as shown in Figure 15. This is because whilst a larger taper angle creates a greater pressure gradient, the fluid is subject to larger upward boundary motion from the channel surface, which tends to slow down the flow going out of the channel. A 20° taper angle appears to reach a sweet spot where these two effects are well balanced for the highest water flow rate.

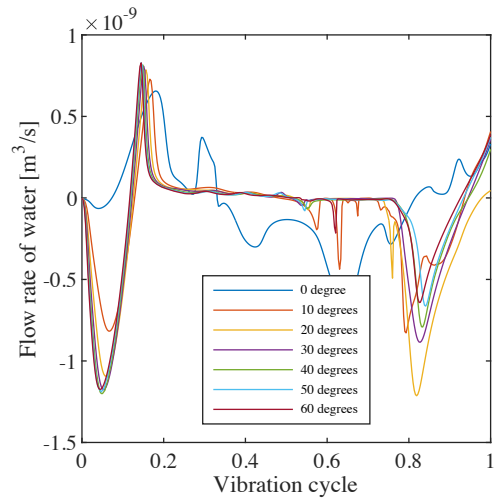

Figure 14 Flow rate of water in the first vibration cycle at different taper angles.

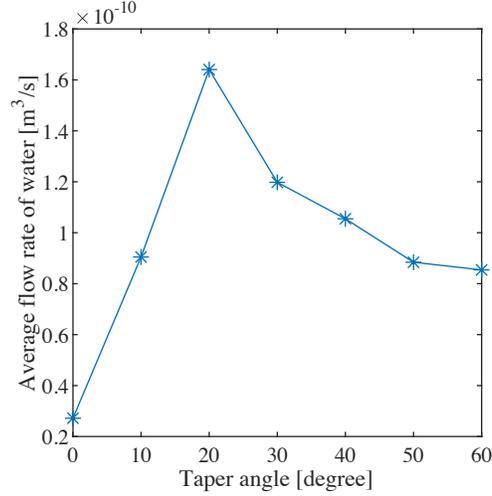

Figure 15 Average flow rate of water at different taper angles. $-Q_1$ is plotted since negative $Q_1$ represents water going out of the channel.

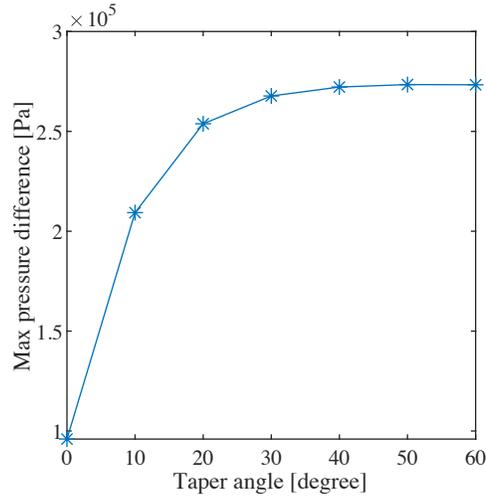

Figure 16 Max pressure difference at different taper angles.

## 4.2 Effects of vibration amplitude and frequency

A harmonic vibration profile is determined by its displacement amplitude and frequency. In Section 3, a linear relationship between the acceleration amplitude and water flow rate is found at 107 kHz, Figure 8, and 29.5 kHz, Figure 11. Given the linear correlation between the displacement ($D$) and acceleration ($A$) amplitude:

$$A = 4\pi^2 f^2 D \tag{21}$$

there should also be a linear relationship between the displacement amplitude and water flow rate with the same degree of correlation ($R^2$). We also evaluate the effect of vibration frequency on the variation of flow rate at the outlet. Five simulations are performed with resonant frequencies (29.5, 41.1, 54.1, 69.6 kHz) lie between the ultrasound lower limit and structural linear deformation upper limit shown in Figure 2. The displacement amplitude is kept at 25 μm. The flow rate time history in the first vibration cycle is plotted in Figure 17, which indicates an increasing trend as the frequency becomes higher. The resultant normalized average flow rate plotted in Figure 18 against normalized frequency exhibits a strong linear relationship with $R^2$ equal to 99.93%, which is higher than that of the displacement amplitude (96.67% as found in Figure 11). The weaker linear correlation between the water flow rate and displacement amplitude may be caused by greater differences brought to the flow behavior as displacement is changed. Since both displacement amplitude and frequency scales linearly with the velocity amplitude ($V$) in harmonic vibration:

$$V = 2\pi f D \tag{22}$$

the water flow rate is linearly proportional to the velocity amplitude. As for the acceleration amplitude, the linear relationship with the flow rate only exists if the frequency is fixed, as found earlier in Figure 8 and Figure 11, since it has a nonlinear relationship with the frequency (Eq. 21).

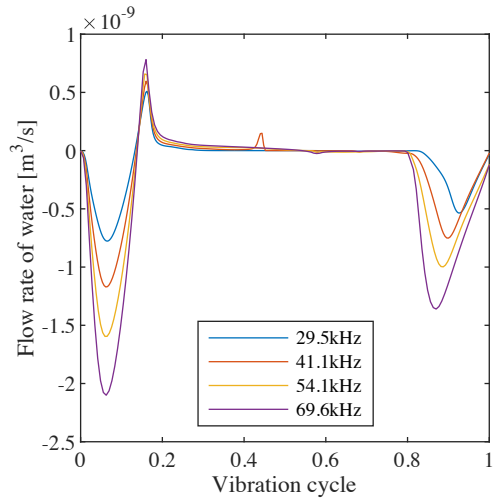

Figure 17 Flow rate of water in the first vibration cycle at different frequencies.

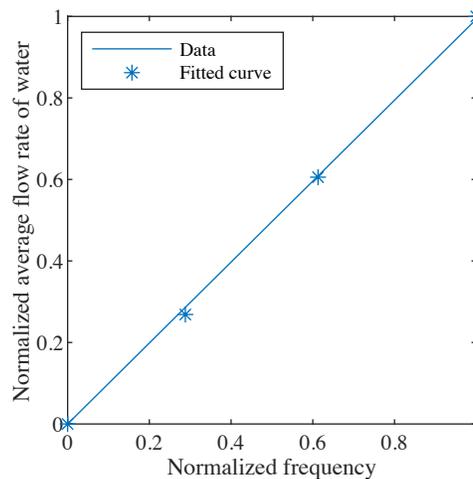

Figure 18 Normalized average flow rate of water at different normalized frequencies. The axis normalization is performed by subtracting the minimum value and dividing by the maximum value to set the range from 0 and 1.

## 4.3 Effects of channel surface contact angle

In the previous investigations, surface contact angle $\theta_w$ is assumed to be 90°, meaning there is no capillary force, a net force along the wall surface caused by the imbalance of the surface tension force, as the contact line moves on the wall. This is rarely practical as a solid surface is usually either hydrophilic ($\theta_w < 90°$) or hydrophobic ($\theta_w > 90°$). Therefore, we assess the effect of contact angle on the water flow rate by additional simulations with a contact angle of 30, 60, 120, and 150°, which covers typical solid surface situations. The frequency is fixed at 29.5 kHz and displacement amplitude at 50 μm. As shown in the water volume fraction contour plot for each contact angle in Figure 19 (at around $\frac{3}{5}$ vibration cycle), the interface contact line moves much slower when the solid surface is more hydrophilic (smaller contact angle) since there exists a larger capillary force pointing towards the air to resist the motion of the phase interface as the air enters the channel. As for the hydrophobic surface, the faster upward-moving contact line, caused by the capillary force pointing towards the water, leads to a delayed water ejection when the pressure gradient is reversed (pointing into the channel) and the water is driven out of the channel, Figure 20 (contour plot at around $\frac{3}{4}$ cycle). The outlet flow rate history in the first vibration cycle shown in Figure 21 consolidates this observation in that the secondary water ejection happens later as the contact angle increases. The average water flow rate is calculated and plotted in Figure 22 with respect to the contact angle, which suggests the highest flow rate is achieved when the contact angle is 60° (slightly hydrophilic). The decrease in the flow rate with the stronger hydrophobicity surface can be explained by the delayed and shortened secondary water ejection within a vibration cycle. It should be noticed, however, the flow rate is not simply inversely proportional to the contact angle as what happens for 30°, Figure 19(a), in which case the water traveling down the channel encounters and get slowed down by the left-over water on the channel surface, Figure 20(a). A 40.8% difference between the highest (at 30° contact angle) and lowest (at 150° contact angle) water flow rate indicates the importance of contact angle (surface wettability) and, more generally, surface tension effect in microchannel flows.

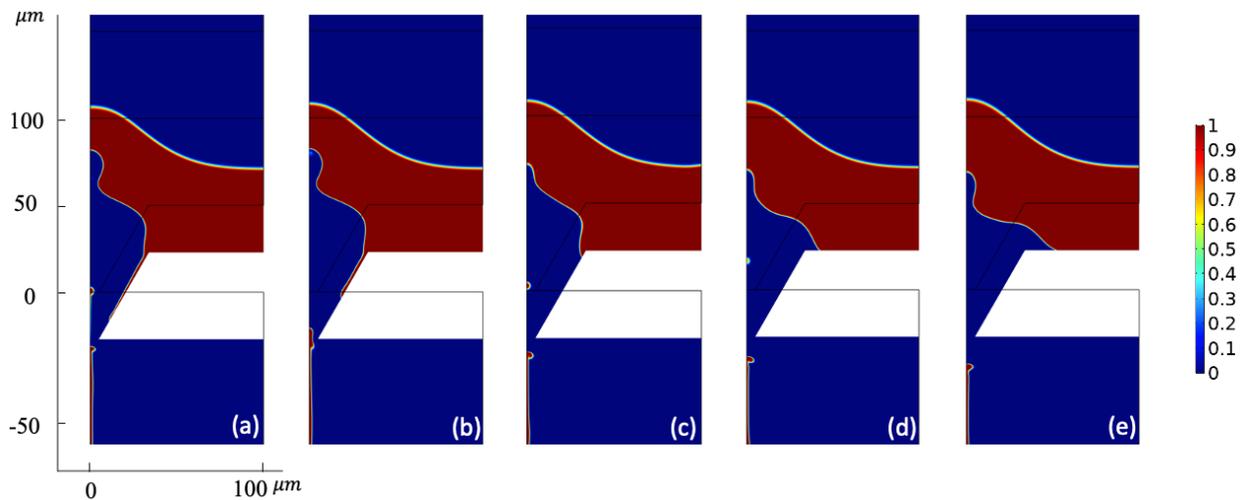

Figure 19 Volume fraction of water contour plot at $\frac{3}{5}$ vibration cycle ((a)~(e): 30, 60, 90, 120, and 150° contact angle).

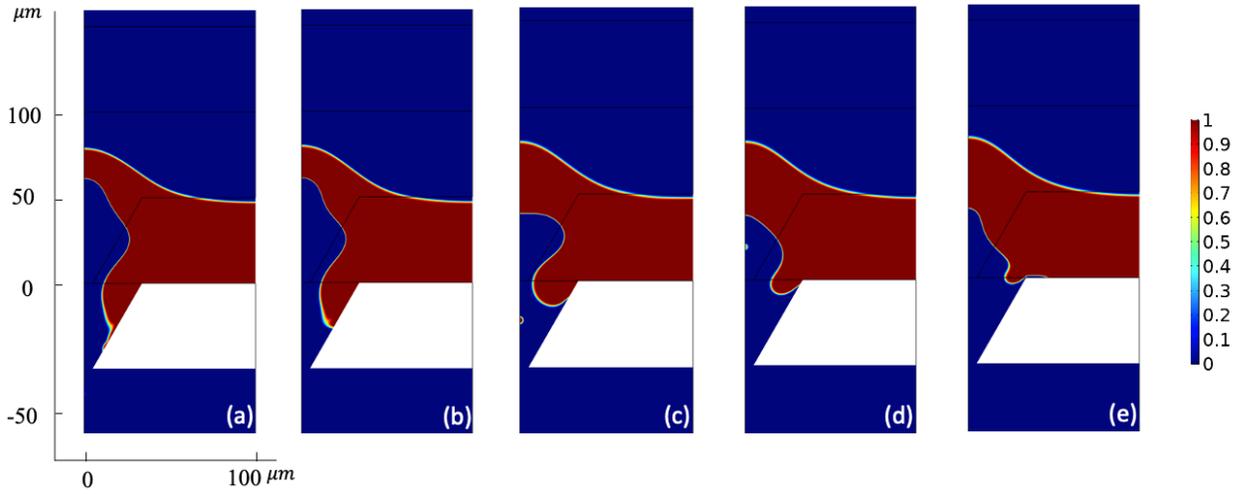

Figure 20 Volume fraction of water contour plot at $\frac{3}{4}$ vibration cycle ((a)~(e): 30, 60, 90, 120, and 150° contact angle).

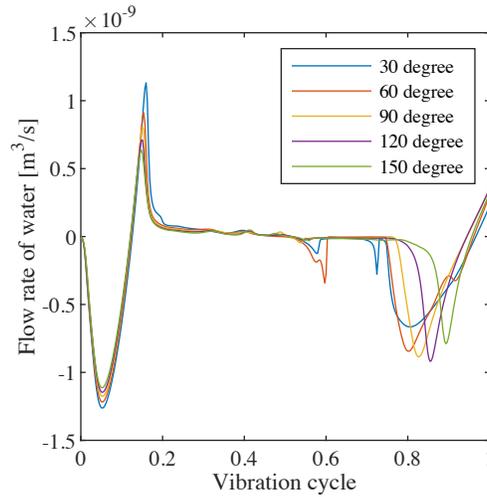

Figure 21 The flow rate of water in the first vibration cycle at different contact angles.

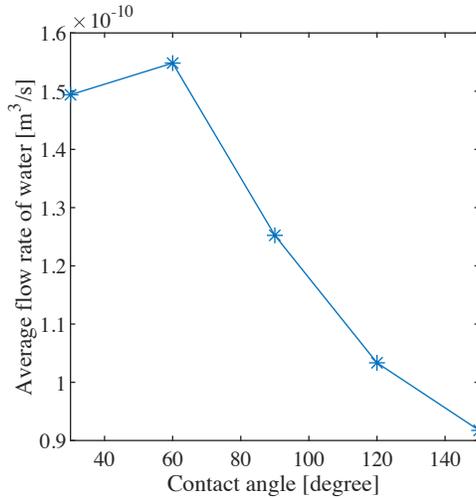

Figure 22 The average flow rate of water at different contact angles. $-Q_1$ is plotted since negative $Q_1$ represents water going out of the channel.

## 5. Conclusions

    This paper offers a glimpse into the ultrasonic oscillatory two-phase flow in the microchannel with the configuration of a novel ultrasonic fabric drying device. Both experimental and numerical investigations are performed with the focus mainly dedicated to searching for accurate CFD modeling techniques and understanding the transport behaviors of the flow. A RANS $k-\omega$ turbulence model coupled with the phase-field method is employed to capture the dynamics of the two-phase flow, which is validated qualitatively by the experimental observations. The flow features a highly unsteady nature involving water ejection through the microchannel outlet and subsequent entrainment of air from the external atmosphere, driven by the alternating pressure gradient during the channel vibration cycle. Atomization is also induced by the free surface faraday instability under the ultrasonic vibration.

    Parametric studies covering the effect of channel taper angle, vibration conditions, and surface contact angle on the channel flow rate are performed with the aid of CFD simulation. It is found that the taper angle that facilitates the highest water flow rate through the outlet is around 20°, although a larger taper angle could create a greater pressure gradient. For the influence of vibration conditions, the flow rate is found to be linearly proportional to displacement amplitude and frequency, which further implicates a linear relationship between the velocity amplitude and flow rate, and a nonlinear relationship for the acceleration amplitude if the frequency is changed. For different channel surface contact angles due to different surface wettability, 60° (slightly hydrophilic) renders the highest flow rate. Hydrophobic ($\theta_w > 90°$) surface facilitates air entrainment and thus causes a delayed and shortened secondary water ejection within a vibration cycle. On the other hand, a too hydrophilic of a surface leads to left-over fluids on the channel surface, which slows down the flow going out of the channel. A significant difference between the flow rate with different surface contact angles reveals the importance of capillary force and, more broadly, surface tension effect in this type of flow.

Although the oscillatory two-phase flow in the microchannel is assessed in the background of ultrasonic drying, the modeling approaches and the results of parametric studies could be easily transferred to other applications involving similar flow behaviors.

## Acknowledgment

This work was sponsored by the US Department of Energy's Building Technologies Office under contract no. DE-AC05-00OR22725 with UT-Battelle, LLC. The authors would also like to acknowledge Mr. Antonio Bouza, Technology Manager—HVAC&R, Water Heating, and Appliances, US Department of Energy Building Technologies Office.

## Data Availability

The data that support the findings of this study are available from the corresponding author upon reasonable request.

## Appendix A: Mesh independence study

As mentioned in Section 2.3, the mesh used in this work is composed of elements of almost uniform size, Figure 23, to prevent excessive diffusion caused by the large capillary width $\epsilon$ constrained by the largest element in the mesh. To achieve an accurate numerical solution, the sharp-interface limit of the Phase Field Method should be reached by reducing the capillary width $\epsilon$, which scales with the diffuse layer thickness. For two-phase flow problems involving moving contact line, the sharp-interface limit is only reached if the mobility parameter $\gamma$ is kept the same while the capillary width $\epsilon$ is reduced, as suggested by Yue[43]. The configuration corresponds to the highest vibration amplitude used to validate the linear relationship between the average water flow rate and the acceleration in Figure 11 is employed to demonstrate how the mesh-independent solutions are obtained in this paper. Totally 5 meshes with total DOF increments by 1.5 times per refinement are tested with the same $\gamma$, which is set as the square of capillary width ($\epsilon^2$) of the finest mesh. The mesh information is listed in Table 1. The independence of the solution generated by different meshes is assessed by two metrics. The first is the L2 norm of the difference of the outlet water flow rate time history in the first vibration cycle (shown in Figure 24) between two consecutive refinements. The second is the average water flow rate calculated for each mesh. The results can be found in Table 1. Since the difference in the average flow rate and the L2 norm between mesh 4 and mesh 5 is decreased to a sufficiently small value, the solution of mesh 4 is considered to be mesh independent.

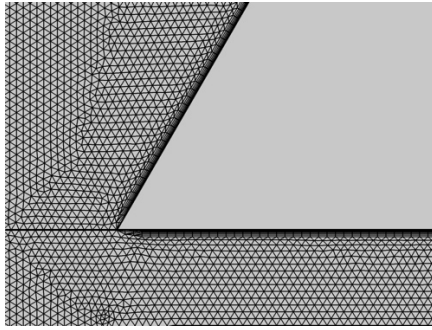

Figure 23 Close-up snapshot of the mesh near the channel outlet (Mesh 4).

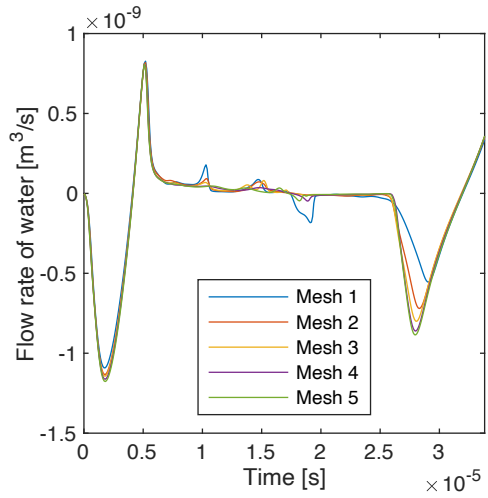

Figure 24 Flow rate time history for different meshes.

Table 1 Information for mesh independence study. Mesh 1 is the coarsest mesh and Mesh 5 is the finest mesh. L2 norm is recorded under the coarser one for two consecutive meshes.

| Mesh | DOF | Average flow rate [$m^3/s$] | L2 norm |
|------|--------|-------------------|----------|
| 1 | 52660 | 8.76e-11 | 7.28e-11 |
| 2 | 76647 | 1.00e-10 | 2.80e-11 |
| 3 | 112492 | 1.07e-10 | 2.17e-11 |
| 4 | 161848 | 1.15e-10 | 1.78e-11 |
| 5 | 235062 | 1.19e-10 | |